# CRACK-LIKE PROCESSES GOVERNING THE ONSET OF FRICTIONAL SLIP

*Shmuel M. Rubinstein, Meni Shay, Gil Cohen, and Jay Fineberg*
*The Racah Institute of Physics, The Hebrew University of Jerusalem, Givat Ram, Jerusalem Israel*
*Author for correspondence (E-mail: jay@vms.huji.ac.il)*

## 1. INTRODUCTION

Although relevant to many aspects of our everyday lives, scientists have yet to fully decipher the fundamental mechanisms of friction – that is, what occurs when two surfaces begin to slide against one another? Frictional slip is central to fields as diverse as physics (Baumberger et al., 1999; Ciliberto and Laroche, 1999; Filippov et al., 2004; Muser et al., 2001; Urbakh et al., 2004) tribology (Bowden and Tabor, 2001; Persson, 2000), the mechanics of earthquakes (Andrews and BenZion, 1997; Ben-Zion, 2001; Bouchon et al., 2001; Cochard and Rice, 2000; Dieterich and Kilgore, 1994; Heaton, 1990; Ohnaka and Shen, 1999; Rice et al., 2001; Scholz, 1998; Xia et al., 2004) and shear-driven fracture (Freund, 1979; Gao et al., 2001; Ravi-Chandar et al., 2000; Rosakis, 2002). Beyond their fundamental scientific interest, questions in the field of friction have relevance to the issue of earthquake measurement and predictions, as well as for numerous industrial applications.

In 1699 Guillaume Amontons gave the modern formulation of the static friction laws. In 1785 Charles-Augustin de Coulomb added the first formulation of dynamic friction laws and suggested a model that partially accounts for them. The explanation as to why a friction coefficient exists and is independent of the (apparent) contact area was introduced in the Micro Contact Interface model (MCI) of Bowden and Tabor (Bowden and Tabor, 2001) in the 1940's. These authors were the first to note that, due to surface roughness, the real (net) area of contact actually consists of a myriad of micro-contacts. The *net* contact area formed by these micro-contacts is only a fraction of the *apparent* contact area. Bowden and Tabor noted that it is the net contact which is responsible for the frictional force. The amount of net contact area formed is dependent on neither the surface roughness nor geometry of the bulk material, but only on the material's mechanical properties and the normal load applied. When two surfaces are brought into contact, their highest asperities touch first, and the load is carried by these contact points. When the pressure at the contact points is larger than the microscopic yield stress of the material, these contact points flow together to form junctions which first elastically deform to increase the net contact area. When, however, the pressure at a micro-contact approaches the material's yield stress, the net contact area will increase due to both plastic deformation and material flow. The resistance of these "welded" junctions to shear is the friction force. Here, we will present the results of direct measurements of the local net contact area along the entire interface in nearly "real" time.

## 2. EXPERIMENTAL METHOD

Before describing our results, we present, in this section, a detailed description of our experimental system. Our experiments describe the dynamics of two brittle PMMA (Polymethylmethacrylate) blocks separated by a rough interface. Each surface, consisting of many discrete randomly distributed micro-contacts, was artificially roughened to an approximate roughness of up to 2μm (rms). Slip is initiated by loading the blocks with a constant normal force $F_N$ while slowly increasing the applied shear force, $F_S$. We measure the dynamics of the interface by means of real-time visualization of the net contact area formed by the rough interface between the two blocks.

A schematic diagram of the experimental apparatus is presented in Figure 1a. Two PMMA blocks, a slider and base, are placed one on top of the other. The respective dimensions of the slider and base are 100-200 x 6 x 75 mm and of size 300 x 27 x 30 mm, where we define $X$ as the direction of applied shear, $Y$ as the sample thickness and $Z$ as the normal loading direction. A rough interface was formed in the following manner. The slider's contact face was first diamond machined to a flatness of better than 0.05μm (rms) and then roughened to degrees of roughness which were varied between 0.05-5μm rms. The base block was first machined to within a 10μm flatness, and then hand-lapped to roughnesses which were varied to between 0.4-10μm rms. A non-machined, commercially flat (50nm rms) base was also used.

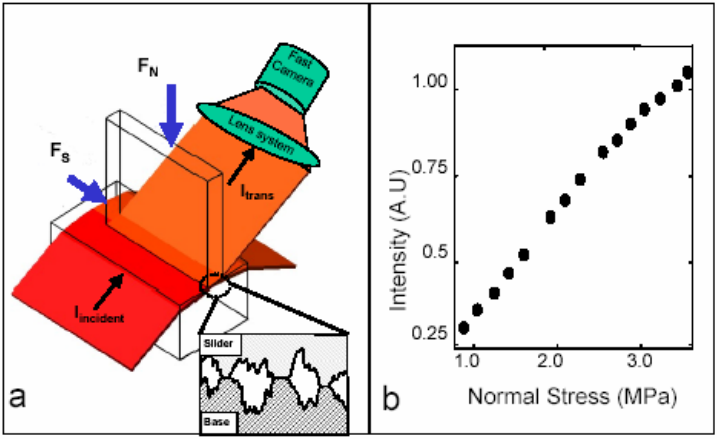

**Figure 1: (a) A schematic view of the experimental apparatus. A laser sheet is incident on a rough interface (inset) at an angle far beyond the angle for total internal reflection from the PMMA-air interface. Light can only traverse the interface at either the actual points of contact or by tunneling across the interface. The transmitted light, imaged on a fast camera,**

is therefore a direct measure of the net contact area at each point. (b) For an interface formed by two rough materials the integrated intensity of light, hence the total net contact area, is proportional to the applied normal stress, $F_N$. The data in the figure were obtained for an interface whose 1μm RMS roughness is much greater than the 50nm tunneling distance at the 70° angle of incidence used.

A normal load, $F_N$, ranging from 1-5000N (1-10MPa depending on both $F_N$ and the surface area of the sample), was applied uniform in $X$ and $Y$ to the slider. This was accomplished by coupling a stiff S-beam load cell, to which a point load was applied, to the slider via a soft "mattress", consisting of an array of 40 soft springs with a total stiffness of 4 x $10^5$ N/m. Although there was no feedback mechanism controlling the application of normal load, we measured $F_N$ fluctuations of less than 2% during shear loading and overall sliding.

The shear load, $F_S$, which was varied from 0-3 MPa, was applied via a stepping motor moving in discrete steps of 0.06μm with loading rates ranging from 1μm/sec to 10mm/sec. The stepper motor applied the shear stress to one side of an S-beam load cell of stiffness of 4 x $10^6$ N/m. The opposite side of the load cell was attached to a 12mm diameter rod, which pressed against the trailing edge of the slider. Thus, shear was applied to the sample over an area whose center of mass was 6.5 mm above the interface. This introduced a small mechanical torque ($2.6 \cdot 10^{-3} \cdot F_N$ N·m) in the loading which resulted in an initial 2.5% variation of $F_N$ over the length of the interface. Our resolution in the dynamic measurement of shear was limited by the compliance of the loading system. For example, the sharp 5-8% decrease of $F_S$ upon the onset of sliding was accompanied by slight damped oscillations whose maximum amplitude was about $0.005 F_S$ and duration about 0.5ms. This limited our measurement accuracy for the trailing edge slip to about 2.5 μm.

The interface was illuminated by a 200 x 5mm laser sheet ($\lambda$=660nm) whose angle of incidence ($\alpha$) was much larger than the critical angle needed for total internal reflection (over 60º) from the acrylic-air interface. Since the cross-section of the laser sheet parallel to the interface plane is 200 x 5/$cos(\alpha)$, the laser sheet illuminated the entire interface 6mm wide. The intensity of both the transmitted and reflected light as a function of space and time was imaged by a fast CMOS sensor (VDS CMC1300 camera). The sensor can be configured to frame sizes of 1280 x $N$ pixels with frame rates of 500,000/(N+1) frames/sec. With this sensor we were able to visualize the instantaneous net contact area along the entire interface at rates of up to 4μsec per frame. For a 150 x 6 mm interface imaged at 100,000 frames/sec (typical conditions in the experiments described below) each pixel is mapped to a 0.1 x 1.5 mm section of the interface, where the higher resolution is in the propagation direction. Thus, in the experiments described below, we do not measure the contact area of individual micro-contacts but, rather, the total contact area of the relatively large (~ 1500) ensemble of micro-contacts encompassed within each pixel. Individual micro-contact dynamics could be visualized with this method but, with the present 1280 pixel resolution of our sensors, at the expense of detecting the overall large-scale interface dynamics.

For the experimental conditions used, light will only effectively traverse the interface, either at points of direct contact or by tunneling through the distance '*h*' separating the two surfaces. When the interface is sufficiently rough, as in the 1μ RMS typical for our dynamic experiments, the distance separating the two surfaces will be much larger than the exponential decay length (d~50nm) of the evanescent light. As a result, the transmitted intensity, $I(x,y)$, at each point along the interface is proportional to the net contact area at that spatial location. This linearity is demonstrated in figure 1b, where the spatially integrated transmitted intensity is shown to be an approximately linear function of the applied normal load. Thus, the total net contact area along the interface is a linear function of the applied normal stress, providing a direct validation of the Bowden and Tabor picture of friction (Bowden and Tabor, 2001; Persson, 2000). As will be described in detail in the next section, the slight curvature in evidence in Fig. 1b is due to small tunneling contributions to the intensity that occur at high values of $F_N$.

Data acquisition of the contact area was performed by the use of a circular buffer of size 128MB into which interface visualization data was continuously acquired. Cessation of the acquisition was triggered at a preset delay from the onset of sliding. The triggering signal was obtained by means of an acoustic sensor coupled to the leading edge of the slider. In this way, we were able to acquire contiguous measurements of the contact area over the time interval of 250ms surrounding the onset of motion.

## 3. CONTACT MODELS AND DESCRIPTION OF THE IMAGING METHOD

Let us first derive the Amontons-Coulomb law in the MCI form (Bowden and Tabor, 2001). Two assumptions are made:

$$F_N = p \cdot \Sigma \tag{1}$$

and

$$F_S = s \cdot \Sigma \tag{2}$$

where $F_N$, $\Sigma$, $F_S$, $p$, $s$ are, respectively, the normal load, real area of contact, peak tangential strength, penetration hardness (which is roughly three times the yield strength (Rabinowicz, 1995)) and shear strength respectively. Equation (1) states that the real area of contact is proportional to the load while in (2) the frictional force is proportional to the real area of contact. By combining (1) and (2) we obtain the static friction coefficient, $\mu_s=s/p$.

Each of the surfaces in contact can be described, in principle, by its topography function $Z(x,y)$, where $Z$ is the height of the surface at the point $(x,y)$. It has been argued (Archard, 1957; Greenwood and Williams, 1966) that, in order to treat friction, we can think of a surface as an ensemble of asperities of spherical shape with a given density, height distribution and radius distribution, designated as n, $\Phi(z)$ and $F(\rho)$ respectively. $\Phi(z)$ can be assumed to be a Gaussian distribution whose width, $\sigma$, is the roughness of the surface. $F(\rho)$ is taken to be a delta function of the asperity radius, $\rho$. This constant average asperity radius approximation does not significantly influence the results of the following MCI model (Buczkowski and Kleiber, 2000; Persson, 2000). Furthermore, it is possible to treat one surface as perfectly flat by adjusting the parameters used to describe the second surface. We follow this statistical approach and numerically generate a topography function $Z(x,y)$ having a constant asperity radius $\rho$, a given asperity density, n and a Gaussian height distribution of width $\sigma = \sqrt{(\sigma_1^2 + \sigma_2^2)}$, where $\sigma_1$ and $\sigma_2$ are the roughnesses of the two surfaces forming the interface. The values of these parameters where determined by surface scans performed with a contact profilometer (Taylor and Hobson Surftronix 3+), and verified with an atomic force microscope. It should be noted that $\sigma$ is measured to high accuracy, while $\rho$ and n are only roughly evaluated from these surface scans. All of the calculations to be presented are performed using computer generated topography functions. The results are independent of the specific surface generated.

The intensity $I(x,y)$, of the transmitted light at each point $(x,y)$ along the interface is given by:

$$I_s(h) = I_0 \int \begin{cases} e^{\frac{-(h-Z(\mathbf{x},y))}{d}} & \text{if } h > Z(\mathbf{x},y) \\ 1 & \text{otherwise (full contact)} \end{cases} \qquad (3)$$

where $d$ is the evanescent decay length defined by:

$$d = \frac{\lambda}{4\pi} \cdot ((n_1 \sin \theta_i)^2 - n_2^2)^{-\frac{1}{2}} \qquad (4)$$

In Equation 3, $h$ is the distance between the mean of $Z(x,y)$ and the flat opposing surface. Full transmission of light only occurs whenever there is contact between the surfaces, otherwise the intensity is determined by the amount of tunneling between the two surfaces.

Let us treat the two extreme cases of either fully plastic or fully elastic deformation at the asperity tip. For fully plastic deformation, the load carried by the interface as a function of the average separation distance $h$ is given by:

$$F_N(h) = \pi \cdot n \cdot A \cdot \rho \cdot Y \cdot \int_h^\infty (z-h) \cdot \Phi(z) \, dz \tag{5}$$

And for the fully elastic case, by:

$$F_N(h) = \frac{4}{3} \cdot n \cdot A \cdot \kappa^{-1} \cdot \rho^{\frac{1}{2}} \int_h^\infty (z-h)^{\frac{3}{2}} \cdot \Phi(z) \, dz \tag{6}$$

Here, $\kappa$ is defined by $\frac{1-\nu_1^2}{E_1} + \frac{1-\nu_2^2}{E_2}$ where $\nu_{1,2}$ and $E_{1,2}$ are, respectively, the poison ratio and Young's modulus of the two materials forming the interface. Y in (5) and (6) is the penetration hardness of the material and A the apparent surface area of the interface. By numerically inverting equations (5) and (6) we calculate the average distance, $h$, between the surfaces and use Eq (3) to compare the light intensity as a function of the applied load to the experiment. Circles in Figure 2 present measurements of the light intensity passing through the interface as a function of the applied normal load for a well polished surface (50nm RMS). The dotted lines show the calculated transmitted intensities for the experimental conditions using Eq. (3) together with Eq. (5). The calculated curves were obtained by using the best fit values for the fully plastic (Eq 5) model. Two fitting parameters where used. The first parameter used was $\rho \cdot Y$. The second parameter was a constant compensation necessary to account for the loss of light due to non evanescent scattering. The need for this correction is especially noticeable for smooth samples where, at low loads, the majority of the light is not transmitted at points of contact but scattered, since, at low loads, the contact size is much less than the 0.588μm wavelength of the incident light. The values of n and σ were obtained from the profilometer surface scan.

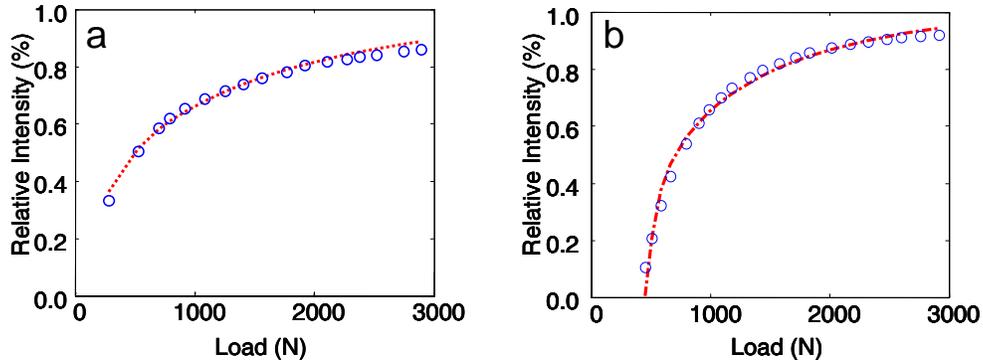

**Figure 2: Relative light intensity transmitted through the interface as a function of the applied load for smooth samples. Measured values (circles) are compared to values calculated using equations 3 and 5 (dashed line). (a) PMMA on PMMA (b) PMMA on Glass In both (a) and (b) both sides of the interface had a 50nm rms roughness.**

The best fit of the plastic model (Fig. 2) to the measured light intensity yields an average stress of 150-400MPa at micro contact tips. This number is consistent with the measured penetration hardness strength by both Briscoe at al (Briscoe et al., 1998) and Dietrich at (Dieterich and Kilgore, 1994) al for similar materials. There is, however a slight but systematic deviation between measured and calculated intensities in Fig. 2. We interpret these slight discrepancies as due to elasto-plastic corrections (Kogut and Etsion, 2002; Kogut and Etsion, 2003) that should be made for the very smooth (50nm rms) surfaces used in Fig. 2.

The fully elastic model (Eq 6) was also tested using $\rho$ as the free parameter. The elastic model produces highly unrealistic values for the fitting parameters. The average asperity radius $\rho \sim 5$nm is over two orders of magnitude lower than the value suggested by the surface scans, and the average stress of 10GPa at the micro contact tip is nearly two orders of magnitude higher than the measured yield stress of PMMA. We thus conclude that, for this system, the pure plastic model of micro-contact tip distortion is the applicable one.

The ratio of elastic to plastic deformation for a given material depends on the quantity $\rho/\sigma$, which is expected to decrease with surface roughness (Persson, 2000). In figure 3 we present the measurements of the transmitted light intensity as a function of the applied load, for the considerably rougher interfaces formed by two PMMA samples with a 500nm rms roughness. Note that, for these surfaces, the fully plastic model provides an excellent fit, yielding both no noticeable systematic deviation and a value of $\rho=0.2\mu$m (Y was set to 200 MPa) in good agreement with the surface scans.

For rough surfaces ($\sigma \gg d$), as presented in Fig. 1b, a linear relation between transmitted light through the interface and normal stress is obtained. This relation is model-independent, as the influence of the different contact models comes in through the degree of tunneling that occurs across the interface, and in the rough interface limit tunneling through the interface is negligible. Thus, these conditions allow quantitative measurement of frictional properties of the interface without the need for any assumptions regarding the specific deformation mechanism.

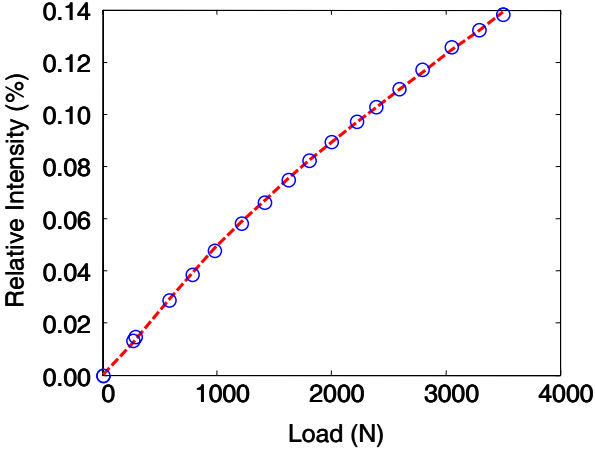

**Figure 3: Relative light intensity (circles) transmitted through the interface as a function of the applied load for rough (500nm RMS) samples. Measurements (circles) are in excellent agreement with values (dashed line) calculated using equations 4 and 5.**

## 4. EXPERIMENTAL RESULTS

### 4.1. Direct Measurements of the Net Contact Area

Slip was initiated by increasing $F_S$ at a constant rate after setting a constant value of $F_N$. The dynamics governing the block detachment process (i.e. interface rupture) are governed by well-defined crack like fronts, which extend in the $Y$ direction and propagate in the $X$ direction along the interface. We use dynamic measurements of $I(x,y,t)$ to study these fronts.

The onset of slip is immediately preceded by the interplay of three different types of coherent crack-like detachment fronts; fast sub-Rayleigh and intersonic fronts, both of which have been previously studied (Rosakis et al., 1999; Xia et al., 2004); and slow fronts (Rubinstein et al., 2004) These fronts propagate across the interface (from the trailing to the leading edge), as shown in Figure 4a and Figure 4b. We define these fronts as "detachment fronts", since they precipitate detachment of the micro-contacts which form the interface. Our measurements are not sensitive enough to determine whether local Mode I type conditions, yielding separation of the interface at the leading edge of these fronts, occurs. Such separation may occur in the dynamics of slip pulses (Ben-Zion, 2001; Bouissou et al., 1998).

Behind a detachment front the net contact area is reduced, whereas ahead the contact area is unchanged. Detachment in our experiments is always initiated by the first of these detachment fronts, the "Sub-Rayleigh" fronts, which propagate at rapid subsonic velocities up to and including the Rayleigh wave speed, $V_R$ ($V_R$=935m/s in PMMA). As seen in Fig. 4c and Fig. 4d, sub-Rayleigh fronts lead to an approximate 10% decrease in the local contact area, immediately upon their passage. These fronts can be identified with earthquakes that propagate in the 0.2-0.8$V_R$ range. Upon detachment initiation, these fronts rapidly accelerate until arriving at velocities in the vicinity of $V_R$. At this point, the sub-Rayleigh fronts suddenly arrest and are *replaced* by two additional types of fronts, which are simultaneously emitted.

The more rapid of these fronts are intersonic fronts, which propagate at speeds which are considerably higher than the shear wave velocity. These types of fronts have been observed in recent experiments (Rosakis et al., 1999; Xia et al., 2004) and calculations (Gao et al., 2001; Needleman, 1999) of explosively induced shear fracture. There is also evidence for such fronts in the seismic records (Bouchon et al., 2001) of recent earthquakes (e.g. Izmit, 1999). Although these waves traverse the entire length of the interface, they give rise to minimal (1-2%) detachment of the net contact area (figures 4c,d). The minimal detachment precipitated by these fronts makes them barely discernable in our direct measurements of net contact area reduction, as shown in figure 4a. Intersonic fronts are, however, clearly visible when the temporal derivative of the contact area at each location along the interface is performed, as demonstrated in figure 4b.

The second type of front, emitted upon arrest of the sub-Rayleigh fronts, is a new type of front, a "slow detachment front". These fronts propagate one to two orders of magnitude more *slowly* than sub-Rayleigh fronts. Our measurements suggest that slow detachment fronts nearly always occur; either as isolated processes or in conjunction with the more rapid modes. These fronts are the most efficient of the fronts for interface detachment, leading to about a 20% reduction in contact area upon their passage. The slow fronts also cause a significant amount of slip while in motion (over 25% of the overall slip before the onset of the overall motion of the slider), but their acoustic signatures are significantly weaker than those of the faster modes. In the experiment presented in Fig. 4a these slow fronts are clearly seen. In this particular experiment, once initiated, the slow fronts traversed the entire interface. In many cases, however, a reverse transition from slow fronts back to sub-Rayleigh fronts occurs (Rubinstein et al., 2004), and the sub-Rayleigh fronts then conclude the interface detachment.

Upon the initiation of sliding, a backward propagating front, labeled as a "rebound front" in figure 4b, is initiated. These fronts, which traverse the entire interface at intersonic velocities, are harbingers of the motion of the leading edge of the sample.

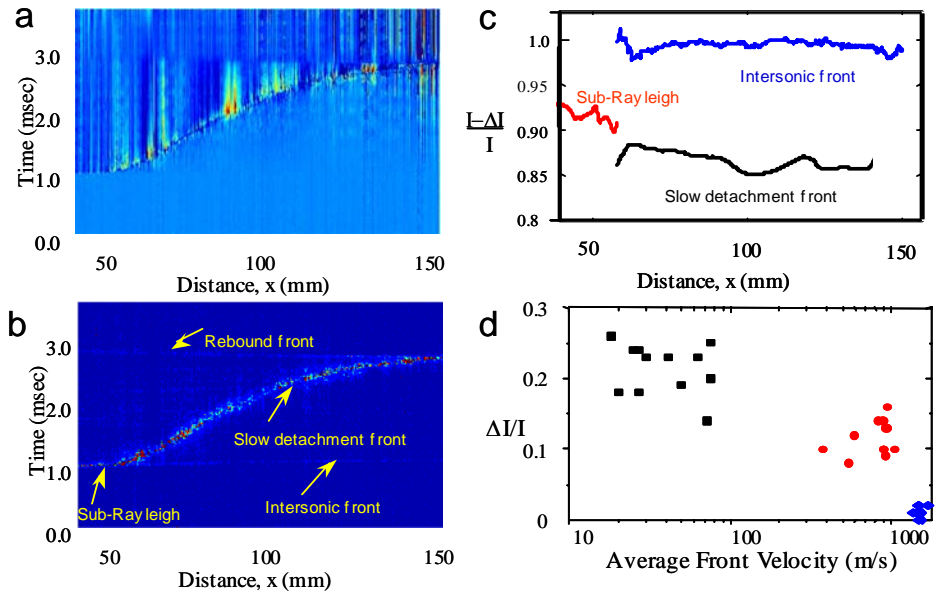

**Figure 4: The dynamics of slip, *prior to overall sliding*, take place via the interplay between 4 different types of coherent crack-like fronts.** (a) Each horizontal line shown is the transmitted intensity $I(x,t)$, averaged over the $y$ direction, of a typical experiment as a function of time. Successive lines spanning the 150mm long interface were acquired at 10μsec intervals. These intensity measurements, normalized by their initial values at each spatial point, are color-coded to reflect the change in the contact area at each spatial point as a function of time. Hot (cold) colors reflect increased (decreased) net contact area. (b) The temporal derivative, $dI/dt$, at each point $x$ along the interface for the data presented in (a), color coded as above. 4 different types of fronts are labeled within the plot. The visibility of the fronts, which can also be seen in (a), is enhanced by the derivation. (c) The relative drop in contact area, as indicated by the intensity drop $(I-\Delta I)/I$ across each of the three forward-propagating fronts shown in (a). (d) The contact area reduction, $\Delta I/I$, in a number of different experiments for slow detachment fronts (squares), sub-Rayleigh fronts (circles) and intersonic fronts (diamonds) as a function of their measured propagation velocities. Note both the large differences in velocities and significant differences in the contact area reductions of the different fronts.

Let us now consider the overall relative motion of the base and slider. We define sliding as the combined motion of both the leading and trailing edges of the sample. Interestingly, sliding does *not* occur as the result of the passage of intersonic fronts, despite the fact that intersonic fronts are seen to traverse the entire interface. Sliding is *only* initiated (Rubinstein et al., 2004) upon the arrival of either sub-Rayleigh or slow detachment fronts at the leading edge of the sample. Slip at the *trailing* edge of the sample is measured while these fronts are in motion. Until their arrival at the leading edge, however, the base and slider remain effectively pinned and no measurable sliding takes place.

Why is no motion at either the leading or trailing edges initiated by the intersonic fronts? A possible explanation is that the intersonic fronts involve only a small sub-population of the micro-contacts. Thus, upon the intersonic front's passage, these "weak" micro-contacts are broken, whereas the larger population of stronger micro-contacts remains intact. This picture provides an explanation for both the negligible amount of contact area freed by these fronts as well as the lack of overall motion precipitated by their passage.

**4b. Photoelastic Measurements**
The total internal reflection imaging method described above is very powerful in supplying real-time quantitative information about the net contact area. However, the data provided by these measurements are limited to the micron thick 2*D* interface. In order to gain insight into the dynamic processes occurring within the elastic bulk above the interface, we conducted additional experiments using the photo-elastic set-up described schematically in figure 5a.

Figure 5b shows a typical photo-elastic pattern visible at a slow frame rate of 1.93 KHz (256 Lines) illuminated with a pulse of 16μs duration. PMMA is not an ideal material for photoelastic tests. Although birefringence can be induced in PMMA, its material fringe constant is almost 20 times higher than polymers usually used in such tests (Ravi-Chandar et al., 2000; Rosakis et al., 1999; Xia et al., 2004). On top of this, our need for spatial resolution in the Z direction limits us to relatively slow frame rates with our present visualization equipment. At these slow rates and long exposure times we are unable to identify rapidly changing individual fringes because of the smearing induced by these effects. The frame rates are rapid enough to capture the dynamics of the slow fronts, but are not adequate for detailed visualization of the sub-Rayleigh or intersonic fronts. Higher frame rates (55,555 frame/sec) have limited resolution in Z (8 lines), but are capable of capturing some of the dynamics precipitated by the more rapid fronts. The qualitative data obtained by this method is indicative of the correlation between the dynamics at the interface plane and the stress distribution within the elastic body to which it is coupled.

All three fronts of detachment are observed in the photoelastic measurements taken in the near vicinity of the interface. Figure 5c shows an (*x*,*t*) plot of the 1*D* stress pattern during the detachment in a typical experiment. The stress is imaged 1mm above the interface (on the slider) at a frame rate of 55,555 frame/sec with a 16μs illumination interval. Note that, unlike the contact area measurements where the intersonic front was responsible only for a mild transient change in the area of contact, the stress change caused by the intersonic front's passage is clearly visible (especially near the transition point) and the change in local stress fluctuations induced by the intersonic front persist at each point, *x*, until the arrival of the slow detachment front. Unlike the work performed in other recent experiments (Ravi-Chandar et al., 2000; Rosakis et al., 1999; Xia et al., 2004), we have neither the temporal nor spatial resolution to resolve the detailed dynamics of the faster fronts by means of photo-elastic measurements. These measurements do establish, however, that the contact area measurements presented above are indeed complementary to previous observations of rapid fronts (Rosakis et al., 1999; Xia et al., 2004). Furthermore, the slow fronts, which have not been previously observed by photo-elastic methods, are easily observable when visualization is performed over sufficiently long temporal intervals.

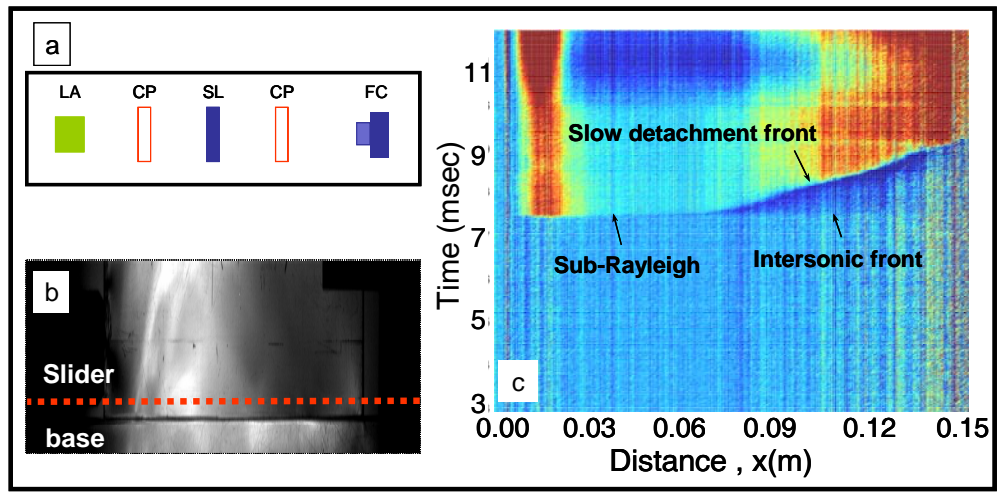

**Figure 5:** *Photoelastic analysis of the dynamics of slip, prior to overall sliding.* **(a) A schematic diagram of our experimental photoelastic setup. A pulsed high intensity, green (505nm), 90 LED array (LA) is circularly polarized (CP) before traversing through the PMMA slider (SL) in the *Y* direction. The light is then passed through an additional circular polarizer (CP) and imaged onto our high-speed digital camera (FC). (b) A typical (256 x 1280 pixel) photoelastic image obtained with a 16μs exposure time. The dashed line marks the location above the interface which was sampled at high (55,555 frame/sec) sampling rates. (c) An (*x*,*t*) plot showing the stress distribution 1mm above the interface (dashed line in b) during the transition to overall sliding in a typical experiment in which $F_N$ = 3500*N* and the shear loading rate was 700*N*/sec. Successive lines spanning the 150mm long sample were acquired at 18μsec**

**intervals. The three different types of forward propagating detachment fronts (labeled in the figure) are clearly visible.**

## 5. Conclusions

The dynamics of friction have, traditionally, been considered to be governed by processes that occur at slow time scales, when the entire slider is in motion. We have studied the detailed processes that occur at the interface between two blocks of like material, during normal loading and immediately prior to the onset of frictional sliding. We first demonstrated how the nominal area of contact at an interface can be measured and imaged. We showed that the real area of contact is proportional to the normal load and we demonstrated that the deformation of the asperities in this material is mainly plastic.

When sufficient shear force is applied to the system, the detachment process, in which extremely short time scales are dominant, takes place via the interplay of 3 different types of detachment fronts that propagate along the interface. We showed that slow detachment fronts, that travel at a speeds one to two orders of magnitude below the shear speed of the material, determine the over all time scale of the detachment process. This time scale is relevant to issues such as the initiation of healing and aging of an interface.

*No* overall sliding of the two blocks occurs until either of the slower two modes traverses the *entire* interface. The onset of sliding, which is initiated by the detachment fronts described above, *is* the transition from static to dynamic friction. The 10-20% reduction in the static coefficient of friction for PMMA is consistent with the overall 12% average reduction in the net contact area that we observe prior to the onset of sliding. Thus, this work elucidates both the mechanisms responsible for the transition from static to dynamic friction as well as possible key processes that take place in dynamics of fault nucleation.

Although entirely reproducible, our quantitative results have been obtained solely for PMMA on PMMA and PMMA on glass interfaces. A relevant question is that of the generality of the observed phenomena. Both the sub-Rayleigh and intersonic fronts have been observed in a series of recent experiments, in which Mode II fracture was studied along weak interfaces (Coker et al., 2005). In these experiments, which were also performed on brittle polymers, entirely different loading configurations, sample sizes and geometries were used. The results of these experiments consistently demonstrated the existence of both sub-Rayleigh and intersonic fronts. These fronts were also observed in both finite element (Needleman, 1999) and molecular dynamic (Abraham, 2003; Gao et al., 2001)simulations which were motivated by these intriguing experimental results.

In more recent experiments by Xia et al (Xia et al., 2004), the transition from sub-Rayleigh to intersonic fracture was studied in detail. Cracks, moving along a weak interface, were observed to propagate at the Rayleigh wave speed, $V_R$, until the shear stress at the shear wave front (traveling ahead of the main rupture) becomes high enough and allows the creation of a secondary daughter crack traveling at intersonic speeds. The transitioning to intersonic occurs at an $F_N$-dependent critical length. This transition process and dependence of the critical length on $F_N$ is consistent with a modification (Xia et al., 2004) to the Burridge-Andrews model (Andrews, 1976). Our experiments yielded no systematic dependence of the transition length with normal stress. Instead, we observed that the generation of intersonic fronts (together with slow fronts) was triggered when a sub-Rayleigh front reached approximately $V_R$. This apparent contradiction may be understood in the following way. A condition of the Burridge-Andrews mechanism for the transition to intersonic fronts is that, when the transition occurs, the initial front is propagating *at* the Rayleigh wave speed. The loading conditions in [(Xia et al., 2004)] indeed produced fronts which, within a few millimeters, attained $V_R$. In our experiments, the acceleration to $V_R$ was much less rapid, and this condition was not met until, in general, the sub-Rayleigh fronts were beyond the transition lengths measured in Xia et al's work. The transition to intersonic fronts at approximately $V_R$ in our experiments may then simply reflect the fact that our fronts were already beyond the transition length at the time when $V_R$ was attained.

    Additional important differences between the two experiments are in the loading configurations and initial conditions. In our experiments a constant value of $F_N$ is first imposed. Shear stress is then adiabatically increased until the transition to sliding (which is mediated by the detachment fronts) occurs. Under these conditions, our experiments are always conducted slightly *above* the shear stress needed to overcome static friction. In the Xia et al. experiments, the system was always *stable* to sliding. These experiments were always conducted for stress levels that were *below* those needed to overcome static friction, but above the stress levels needed to maintain dynamic friction. Motion was triggered in the Xia et al. experiments by the plasma-induced pressure caused by exploding a wire imbedded within the interface. This is not a small perturbation, and its effect is decidedly nonlocal. The rupture fronts along the interface were triggered by the propagating shear wave generated by the explosion. We surmise that the shear wave was necessary to induce the initial motion along the interface, which locally "boosted" the system beyond the static friction threshold. This initial motion then could evolve into an intersonic rupture. As this perturbation propagated at the shear wave velocity, intersonic rupture was, necessarily, initiated ahead of the sub-Rayleigh fronts. As in the two experiments, both the initial perturbations and loading conditions were so cardinally different, it is not entirely surprising that the transition mechanisms between sub-Rayleigh and intersonic fronts would be substantially different.

The dissimilarities in the dynamic interplay between the sub-Rayleigh and intersonic fronts are, perhaps, related to the differences in the triggering and loading conditions in the two experiments. In the Xia et al. experiments, sub-Rayleigh and intersonic fronts were observed to *simultaneously* propagate along the interface, with the initiation of the intersonic fronts occurring at a finite distance ahead of the sub-Rayleigh front. In our experiments, intersonic fronts were only generated upon *arrest* of the sub-Rayleigh fronts and, to our resolution, were initiated at the point of arrest. As the Burridge-Andrews mechanism predicts simultaneous propagation of both types of front, this observation provides additional evidence that this mechanism may not be the relevant one in our experiments.

The character of the intersonic waves in both experiments also appears quite different. Whereas the experiments in (Coker et al., 2003; Rosakis et al., 1999; Xia et al., 2004) indicate that finite displacement of the interface occurs in the wake of the intersonic fronts, in our experiments the amount of slip displacement generated by the intersonic fronts is negligible. The negligible contact area reduction precipitated by intersonic fronts was the basis for our hypothesis that only a small sub-population of the asperities takes part in their motion. This hypothesis could also explain their lack of slip displacement in our experiments, as the "unbroken" asperities surrounding those ruptured by the intersonic front would still pin the interface, allow only negligible slip.

Slow fronts were not reported in the previously described experiments (Coker et al., 2003; Xia et al., 2004), although both sub-Rayleigh and intersonic fronts were readily observed. The reason for this is probably due to the time scales involved. As the high-speed photography used in this previous work had a 50μs maximal duration, the ~1mm that slow fronts would have traveled within this interval would have been nearly undetectable. Given the sound speeds of PMMA ($V_p$=2000m/s, $V_s$=1000m/s) and the geometry of our experimental configuration, reflected waves could play a role in the dynamics of the slow fronts, since time scales are certainly long enough for acoustic waves to traverse the sample numerous times. It is also conceivable that reflected stress waves could affect the dynamics of the fast fronts. Both of these issues require further study.

There is, however, support for the generality of the slow fronts from two other sources; experiments performed by Ohnaka (Ohnaka, 2003; Ohnaka and Shen, 1999) on granite-granite interfaces and evidence obtained from the study of earthquakes. Ohnaka, using a discrete array of sensors distributed along a granite interface, reported evidence of coherent disturbances moving along the interface at much slower velocities than $V_R$. Interpreted as "nucleation zones", these disturbances preceded rapid sub-Rayleigh propagation and their propagation velocity was dependent on the interface roughness. For roughness values comparable to our experiments, these nucleation zones expanded at scaled velocities (~0.01$V_R$) comparable to those that we obtain for slow fronts in PMMA.

Further support for the existence of slow fronts in nature is given by recent reports in the earthquake literature of "slow" or "silent" earthquakes. These observations (Crescentini et al., 1999; Linde and Sacks, 2002; Rogers and Dragert, 2003) demonstrated the existence of events in which significant slip occurred, but with a minimal acoustic signature. These events propagated at velocities significantly slower than $V_R$ for hundreds of kilometers. Such slow events also accompanied (Ammon et al., 2005; Bilham, 2005; Lay et al., 2005) the recent Sumatra-Andaman earthquake of magnitude 9.3. Both analysis and measurements of the slip in this massive earthquake indicated that an a-seismic slow front propagated at approximately 300m/s (less than 0.1$V_R$) for nearly an hour in the northern section of the rupture and accounted for about 1/3 of the total slip in this devastating event.

As this series of experiments appears to successfully model the rapid sub-Rayleigh and intersonic events that are associated with earthquakes, it is conceivable that the slow fronts may indeed be analogs of these recently observed "silent" earthquakes. Little is currently known about such slow earthquakes, but evidence is mounting that these events may be much more common than was originally perceived. In our experiments slow fronts are nearly always observed. Simulations(Ben-Zion and Andrews, 1998; Lapusta and Rice, 2003; Lapusta et al., 2000) of earthquake nucleation suggest that slow nucleation fronts occur in the initial period of earthquake formation. If slow detachment fronts are indeed analogous to this type of earthquake-generated motion, it may be possible to gain significant insight into these important but elusive events by modeling them in the laboratory.